\documentclass[12pt,a4paper]{iopart}
\usepackage{graphicx}
\usepackage{amssymb}
\usepackage{url} 

\begin{document}

\title[Suggested design of THz source based on gold nanoparticles]
{Suggested design of gold-nanoobjects-based
terahertz radiation source for biomedical research}
\author{A.V. Postnikov$^1$ and K.A. Moldosanov$^2$}
\address{
$^1$Universit\'e de Lorraine, LCP-A2MC, 1 Bd Arago, F-57078 Metz, France \\
$^2$Kyrgyz-Russian Slavic University, 44 Kiyevskaya st.,
Bishkek 720000, Kyrgyzstan}
\ead{andrei.postnikov@univ-lorraine.fr}

\vspace{10pt}
\begin{indented}
\item[]January 2018
\end{indented}

\begin{abstract}
Gold nanoparticles (GNPs) may serve as ``devices'' to emit 
electromagnetic radiation in the terahertz (THz) range,
whereby the energy is delivered by radio frequency
or microwave photons which won't by themselves induce transitions between
sparse confinement-shaped electron levels of a GNP, but may borrow the energy 
from longitudinal acoustic phonons to overcome the confinement gap.
Upon excitation, the Fermi electron cannot relax otherwise than via emitting
a THz photon, the other relaxation channels being blocked by force of shape
and size considerations. Within this general scope that has been already
outlined earlier, the present work specifically discusses two-phonon processes, namely
(i) a combined absorption-emission of two phonons
from the top of the longitudinal acoustic branch, and (ii) an absorption of
two such phonons with nearly identical wavevectors. The case (i) may serve                                                                                                                                                                               as a source of ``soft'' THz radiation (at $\simeq$0.54~THz), the case (ii) 
the ``hard'' THz radiation at 8.7~THz. Numerical estimates are done for crystalline
particles in the shape of rhombicuboctahedra, of 5 -- 7 nm ``diameter''.
A technical realisation of this idea is briefly discussed, assuming the deposition
of GNPs onto / within the substrate of Teflon\textsuperscript{\textregistered}, 
the material sustaining high temperatures and transparent in the THz range.
\end{abstract}


\section{Introduction}
\label{sec:Intro}
The elaboration of simple and practical sources of terahertz (THz) radiation,
in spite of a number of solutions presently available \cite{Proc2004FEL-216},
remains an important task in biological and medical research.
In particular, generation techniques have been suggested whose essential element
are metal films \cite{OptLett29-2674,PRL98-026803,OptLett39-777}.
In a recent work \cite{BeilJNano7-983}
we argued that the gold nanoparticles (GNPs) of special size and shape (nanobars
or nanorings) may be used, under special conditions, for generating
radiation at $\sim$4~THz. Later on \cite{Ferroel509-158},
we suggested a scheme of turning an array of GNPs into a working unit 
of a THz to infrared image converter, that would make possible a visualisation 
of pathological tissues which show contrast with the normal tissue notably
in the THz range. 
We made use of the fact that longitudinal acoustic (LA) phonons have frequencies in the THz range,
notably yielding (in bulk \cite{PRB87-014301}, or nanoparticles \cite{Nanoscale6-9157}
of gold) the major peak in the density of modes at 3.9 -- 4.6~THz.
The phonons may be absorbed or emitted by the electron system, whose energy levels
are quantified due to the nanoparticle's small size. (In the following discussion,
the energy steps between which the electron transitions may occur are of the order
of THz photon energy).  Moreover, the external electromagnetic field at different frequencies
may intervene, to deliver energy to the system. Depending on the relation
between respective frequencies and energy intervals, the following scenarios have previously
attracted our attention:

(a) Excitation by electromagnetic radiation at radio frequency (RF; say at 13.56~MHz,
i.e., with the quanta energy by far inferior to the steps in the electron system); 
the excitation of a Fermi electron 
borrows energy from a LA phonon; a subsequent electron's relaxation returns the energy
(with surplus) to phonons; the particle get heated. The efficiency of process
is optimal for GNPs being ``round'' with diameter $\sim$5~nm 
(consistently with experimental findings of Ref.~\cite{NanoRes2-400}; 
see Ref.~\cite{FANEM2015} for details of our theoretical interpretation).

(b) Excitation of electron system by microwave radiation at 2.45~GHz,
assisted by absorption of LA phonons, hindering however the relaxation channel
into emitting a LA phonon by a specific choice of the particles' size / shape, 
namely as nanobars or nanorings, of the length ${\sim}100$~nm and the transversal
diameter ${\sim}3$~nm. As a result, a large part of the excitation energy will be
deviated into a direct emission at THz frequency. The justification and technical
details can be found in Refs.~\cite{BeilJNano7-983,Patent-RU2622093}.

In the present work, we pay attention that in the case (a), once the phonon ``bath''
is hot enough, conditions may appear for two-phonon interaction with the electron system.
Specifically, two processes are imaginable: 

(c) An absorption of a LA phonon along with an emission of a less energetic one
as a single act, the surplus energy being brought away with a ``soft'' THz photon;

(d) A simultaneous absorption of two LA phonons, consequently an excitation of
electron to a much higher level than in a single-phonon absorption, with a subsequent
relaxation via emission of a ``hard'' THz phonon.  

Keeping in mind that the both phonons involved would typically ``belong''
to the major peak in the density of modes of gold, with abovementioned central
frequency and width, the ``mean'' frequencies of the THz photons emitted in the processes
(c) and (d) can be estimated as, say, $\sim$0.54~THz, i.e., with the photon energy $\sim$2.23~meV,
for ``soft'' and $\sim$8.7~THz / 36~meV for ``hard'' cases, correspondingly.
Each of these frequencies is of specific interest for practical use.
The ``soft'' radiation seems to be promising for good contrast in biological scanning:
after \cite{JPhysD39-R301} (a review; cf. in particular Sec.~4.1.4 therein)
and Ref.~\cite{ApplSpectr60-1127}, the maximum difference in refractive index
between diseased and healthy tissue lies between 0.35 -- 0.55~THz, 
whereas the maximum difference in absorption occurs at 0.5 THz 
(cf. in particular Figs.~4 and 6 of the latter publication).
The ``hard'' radiation, in its turn, would offer a better
spatial resolution of the pathological / normal tissue boundaries.

The present work offers a simpe theory analysis of the above cases (c) and (d),
related to electron/two-phonon interactions, taking into account free-electron model
for the metal nanoparticle, the confinement conditions, and the energy / momentum
conservation laws. Differently from the spherical geometry of particles assumed
in our earlier works, we proceeded here from GNPs having the shape of 
rhombicuboctahedra. Such shapes seem realistic in the production of GNPs;
notably, the technology of their preparation with sizes of $\sim$40~nm is outlined
in Refs.~\cite{CrystEngComm12-116,ChemEurJ17-9746}. The particles of smaller sizes
($\sim$5--7~nm), argued to be needed for realisation of the mechanisms we outline,
could be hopefully produced in near future by combination of selective etching and growth reactions.

\section{Particle shape and confinement-imposed quantisation}
\label{sec:shape}
\begin{figure} 
\flushright{\includegraphics[width=0.90\textwidth]{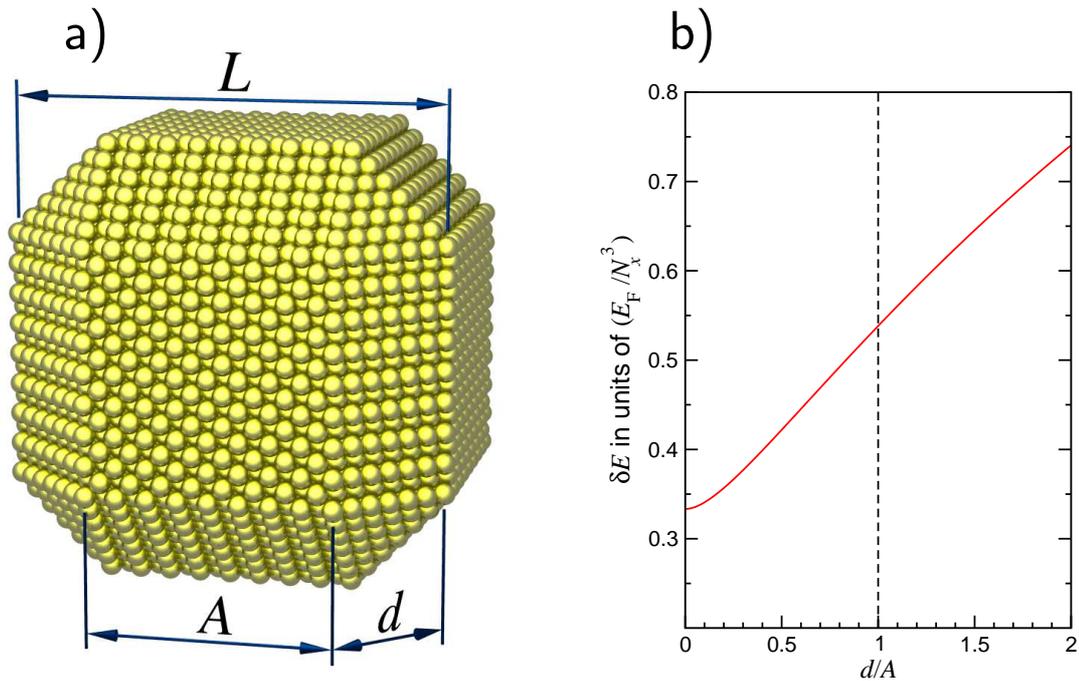}}
\smallskip
\caption{\label{fig:polyhed}
Rhombicuboctahedric gold nanocrystal discussed as a prototype GNP:
a) general view and indicated $A$, $d$ dimensions;
b) The electron energy quantisation step $\delta{E}$ throughout 
different aspect ratio values $d/A$.}
\end{figure}

Out of GNP shapes relevant for practical synthesis,
we consider GNP in the shape of rhombicuboctahedron (Fig.~\ref{fig:polyhed}a)
as a realistic prototype, which comes about in the process of growing / etching
towards cubic or octahedral nanocrystals \cite{CrystEngComm12-116,ChemEurJ17-9746}.
For the edge size $A$ and facet width $d$, 
the ``diameter'' (maximal wall-to-wall distance) of such particle is
$L=A+d\sqrt{2}$, and the volume $V=A^3+3\sqrt{2}A^2d+3Ad^2+(\sqrt{2}/3)d^3$.
For order-of-magnitude estimates of the confinement conditions, we consider first 
the ``ideal'' geometry $d\!=\!A$ of the perfect \emph{small rhombicuboctahedron}
(which has however no special standing from the point of view of
the GNP growth process), whose volume is $V=4(1\!+\!5\sqrt{2}/6)A^3\,\approx\,8.714\,A^3$,
and then discuss the possible effect of varyable aspect ratio $d/A$.

Among the particle's vibration modes 
there will be those propagating at, or under, the surface along closed trajectories;
their energies will be very densely quantified, so that they will be able to absorb 
any surplus energy resulting from the electron relaxation, leading to the GNP's heating.
The ``wall-to-wall'' vibration modes, on the contrary, will be relatively sparsely
discretized.
To be specific, we assume 
the ``wall-to-wall'' distance $L$ to encompass an integer number $N_x$ of gold
lattice parameters $a_{\rm Au}=0.408$~nm; $L=N_xa_{\rm Au}$. 
In the following, we'll refer to phonon dispersion along high-symmetry lines
in the Brillouin zone (BZ) of gold, assuming the nanoparticle for simplicity
to be monocrystalline as shown in Fig.~\ref{fig:polyhed}.
In particular, the wave number measured along the $\Gamma$-$X$ line 
(that is, of the length $2\pi/a_{\rm Au}$, parallel to the principal
cubic axis in the fcc lattice of gold) will be consequently discretized 
with the step ${\delta}q=2\pi/(N_xa_{\rm Au})$.
For comparison, the solution for longitudinal vibrations in free 
uniform beam -- see, e.g., Ref.~\cite{Beards-Chap4} -- yields frequencies
$\omega_n=\frac{n\pi}{L}\sqrt{\frac{E}{\rho}}$
($E$: modulus under an axial force, $\rho$: density) linear (scaled by the speed
of sound) to the wave number $q_n=\frac{n\pi}{L}$, whereby $n$ runs from 0 
to the number of atom layers along the beam length $L$,
hence $q_{\rm max}=\frac{2\pi}{a_{\rm Au}}$ and 
$\omega_{\rm max}=q_{\rm max}\sqrt{\frac{E}{\rho}}$.

The electron states, in their turn, are discretized with the energy step
${\delta}E\,{\approx}\,(4/3)E_{\rm F}/N$, according to the Kubo's formula
\cite{JPSJ17-975,JPhysColloq38-C2-69},
$E_{\rm F}\,{\simeq}\,5.53$~{eV} being the Fermi energy of gold (as immediately follows
from the free-electron model of the corresponding electron density;
otherwise cf.~Ref.~\cite{AshMerm_book}), and 
$N$ the number of gold atoms in a particle of volume $V$. 
In general, 
$N=V/V_{\rm at.}$ with $V_{\rm at.}=a^3_{\rm Au}/4$
(volume per atom in the fcc lattice), and 
$N=16 N_x^3(1\!+\!5\sqrt{2}/6)/(1\!+\!\sqrt{2})^3$ 
in the case of (perfect) small rhombicuboctahedron, so that
\begin{equation}
{\delta}E =\frac{7\!+\!5\sqrt{2}}{6\!+\!5\sqrt{2}}\,{\cdot}\,\frac{E_{\rm F}}{2N_x^3}
\approx 0.538\,E_{\rm F}/N_x^3\,.
\label{eq:delta-E}
\end{equation}
A deviation from the ``ideal'' aspect ratio $d/A=1$ affects the expression for the
GNP volume; the resulting  
quantisation step, for the given particle size $L$ (hence $N_x$)
varies as shown in Fig.~\ref{fig:polyhed}b.
For subsequent quantitative estimates one can reasonably assume that
these poorly controllable variations of the aspect ratio,
combined with imperfections of the particle shape,
relaxations / rounding up at the edges etc. would ``contaminate''
the quantisation step given by Eq.~(\ref{eq:delta-E}) to within some ${\pm}25\%$. 
The Fermi electrons, in the course of their interaction with phonons,
may be in principle excited through an integer number of such steps; 
we'll consider the lowest excitation only.
In the two-phonon excitations outlined as processes (c) and (d)
in Sec.~\ref{sec:Intro}, possible matching conditions need to respect both
the momentum and the energy conservation laws. It is important hereby that
the momenta of phonons may change by uniform steps of ${\delta}(\hbar\mathbf{q})$, 
whereas the energy conservation must take into account the nonlinearity of 
the ``genuine'' phonon dispersion law and ``saturating'' $\omega(q)$ slope
towards the BZ edge. 
Numerical estimations are done in the following sections.  

\begin{figure}[h] 
\flushright{\includegraphics[width=0.9\textwidth]{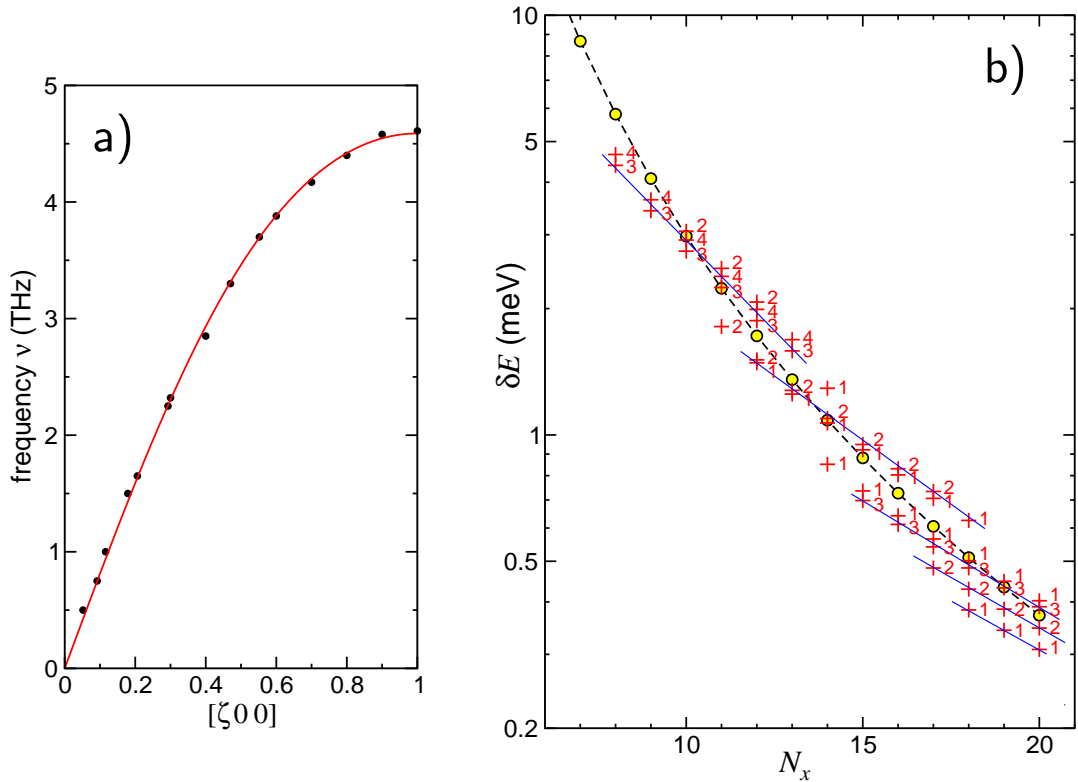}}
\caption{\label{fig:phonon}
Phonon dispersion and its relation to electron level quantisation 
in gold nanoparticles: a) $\Gamma$-$X$ phonon dispersion in fcc Au
from inelastic neutron scattering \cite{PRB8-3493} (dots) and its 
least-square fit after Eq.~\ref{eq:ha-nu}; 
b) ${\delta}E$ as function of GNP size $N_x$ after Eq.~\ref{eq:delta-E}
(connected yellow dots connected by dashed line); red crosses indicate allowed 
$\hbar\omega(\mathbf{q}_1)-\hbar\omega(\mathbf{q}_2)$ values for different $N_x$,
nearly matching ${\delta}E$.
Beside each cross, the number of quantisation steps $d$ is indicated
for the transition in question; see text for details.
Thin descending lines are added as guide to eye to mark
identical quantisation steps throughout different $N_x$ values.}
\end{figure} 

The phonon dispersion in GNP is likely to resemble that of bulk gold,
as is implied by the comparison, shown in Fig.~2 of Ref.~\cite{BeilJNano7-983},
of respective densities of modes for nanocrystals \cite{Nanoscale6-9157} and 
for single-phase ${\sim}10$~$\mu$m thick ribbons \cite{PRB87-014301}.
To numerically characterize the phonon dispersion, we refer
to the inelastic neutron scattering data in Table~1 of Lynn \emph{et al.} \cite{PRB8-3493},
and approximate the dispersion data along $\mathbf{q}=(2\pi/a_{\rm Au})[\zeta\,0\,0]$
by a least-square fit
\begin{equation}
h{\nu}(\zeta) = W_1\sin(\pi\zeta/2)+W_2\sin(3\pi\zeta/2)\,,
\label{eq:ha-nu}
\end{equation}
with $W_1=19.6088$~meV (4.74139~THz); $W_2=0.6356$~meV (0.15368~THz).\footnote{
Here and in the following, the notation $h\nu$ is used along with $\hbar\omega$
in order to justify / facilitate reference to experimental data in THz.}
This fit along with the experimental values of Ref.~\cite{PRB8-3493} is shown
in Fig.~\ref{fig:phonon}a.

\section{Case of phonon absorption - emission}
\label{sec:2ph-abso-emis}
For the lowest excitation of a Fermi electron, 
${\delta}E$ of Eq.~(\ref{eq:delta-E}) should match
the energy difference for absorbed $\hbar\omega(\mathbf{q}_1)$
and emitted $\hbar\omega(\mathbf{q}_2)$ phonons, whereby $\mathbf{q}_1$
and $\mathbf{q}_2$ are discretized by force of particle size confinement.
For numerical estimates, we consider the phonon momenta to be collinear,
in the spirit of the above picture of the ``wall-to-wall'' longitudinal phonon
``resonating'' within the polyhedral particle. Then, the allowed values of $\zeta$
are, for a given $N_x$, identified by the step number $s\,({\geq}0)$ of the incoming phonon,
measured from the BZ border, and by the number of steps $d\,({\geq}1)$ 
between the momenta of two phonons with ``reduced wavenumbers''
$\zeta_1=1\!-\!s/N_x\,;\quad \zeta_2=1\!-\!(s+d)/N_x\,.$
The matching condition 
\begin{equation}
{\delta}E=h\nu(\zeta_1)\!-\!h\nu(\zeta_2)\,, 
\label{eq:matching}
\end{equation}
that unifies Eqs.~(\ref{eq:delta-E})
and (\ref{eq:ha-nu}), works like an equation on $N_x$ for different trial values
of $s$ and $d$. The ``graphic solution'' of this equation is given in Fig.~\ref{fig:phonon}b.
Whereas ${\delta}E$ gradually decreases as function of $N_x$, a number of intervals 
$h\nu(\zeta_1)\!\,-\,\!h\nu(\zeta_2)$ that would nearly match ${\delta}E$
can be found, with different choices of $s$ and $d$.
The ``matches'' which fall within ${\pm}25\%$ of the electronic ${\delta}E$ are marked
in Fig.~\ref{fig:phonon}b by 
red
crosses. Besides each mark, its corresponding $d$ value 
(the number of quantisation steps by which the phonon wavenumber changes)
is indicated; the $s$ number (the placement of the absorbed phonon on the dispersion
branch) is not of immediate importance and hence omitted. For the clarity of this figure, 
only those matches $(\zeta_1,\zeta_2)$ are retained for which both the absorbed and 
the emitted phonons are among ``the most numerous'' ones, i.e. those within the half-width
of the LA peak in the density of modes -- or, differently formulated, those with frequencies
within the upper 30\% of the dispersion branch.

Understandably, for small $N_x$ values and hence sparse steps in the phonon dispersion,
no matches can be found; from $N_x=10$ on, more or less close matches start to appear. 
The ``solutions'' marked by the same relative ``displacement'' of the phonon
momentum, i.e., by the same $d$ value, are placed along regular lines, indicated by thin 
blue
descending lines in Fig.~\ref{fig:phonon}b. Keeping in mind that this graphic solution
can, at most, be considered for order-of-magnitude estimates, we indicate nevertheless
that fairly good matchings according to Eq.~(\ref{eq:matching}) occur for 
$N_x$=10, 11, and then 14 and from 18 on. 

\begin{figure} 
\flushright{\includegraphics[width=0.5\textwidth]{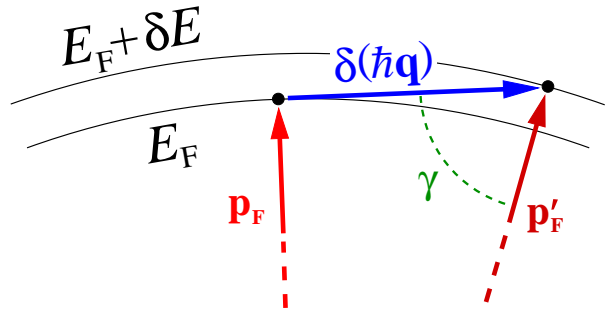}}
\caption{\label{fig:scatter}Scheme of scattering condition
for an electron at the Fermi surface (energy $E_{\rm F}$, momentum $\mathbf{p}_{\rm F}$)
to be promoted to an excited state (energy $E_{\rm F}+{\delta}E$, 
momentum $\mathbf{p}^\prime_{\rm F}$) in the course of absorption / emission of phonons
with residual momentum $\delta(\hbar\mathbf{q})$. See text for details.}
\end{figure} 

Apart from the energy matching, one should consider the momentum conservation
at the phonon scattering. An important observation is that the phonon momenta difference,
$\hbar\mathbf{q}_1\!-\!\hbar\mathbf{q}_2$, must stand almost at normal to the
Fermi electron momentum. Indeed, 
$p_{\rm F}\,$=$\,(12\pi^2)^{1/3}\hbar/a_{\rm Au}\,\approx\,1.27{\cdot}10^{-19}
\,\mbox{g$\cdot$cm/s}$,
and 
\begin{equation}
\delta p_{\rm F}\;{\simeq}\;\frac{p_{\rm F}}{2}\,\frac{\delta E_{\rm F}}{E_{\rm F}}
= \frac{0.538}{2N_x^3}\,p_{\rm F}\,,
\label{eq:delta-pF}
\end{equation}
so (for ``minimalistic'' choice $N_x\,\simeq\,10$) 
$\delta p_{\rm F}\,\simeq\,3.41{\cdot}10^{-23}\,\mbox{g$\cdot$cm/s}$,
whereas the step in the momentum value of phonon is
\begin{equation}
\delta(\hbar q)=\frac{h}{N_x\,a_{\rm Au}}\,{\simeq}\,1.62{\cdot}10^{-20}\,\mbox{g$\cdot$cm/s}\,,
\label{eq:delta-hq}
\end{equation}
two orders of magnitude larger than $\delta p_{\rm F}$. A schematic placement of
the momentum vectors of the Fermi electron prior to, and upon the absorption
of the residual phonon momentum $\delta(\hbar\mathbf{q})$ is depicted in Fig.~\ref{fig:scatter};
a numerical estimate of the angle between the electron and phonon momenta follows from
\begin{eqnarray}
p^2_{\rm F} &=& (p'_{\rm F})^2+[\delta(\hbar{q})]^2-2\,p'_{\rm F}\,\delta(\hbar{q})\cos\gamma\,;
\nonumber \\
\gamma &{\approx}& \arccos\left[\frac{\delta p_{\rm F}}{\delta(\hbar{q})} +
\frac{\delta(\hbar{q})}{2p_{\rm F}}\right] \approx 86^{\circ}\,,
\end{eqnarray}
with the above estimates for the momenta values.
The momentum value $p_{\rm ph}$ of a THz photon that carries away the ${\delta}E$ energy is 
$p_{\rm ph}={\delta}E_{\rm F}/c\,{\simeq}\,1.6{\cdot}10^{-25}\,\mbox{g$\cdot$cm/s}$,
so it can be neglected in the above balance.

Since the ``matching conditions'' in Fig.~\ref{fig:phonon}b can be, in principle,
satisfied for a large variety of ($N_x$,$s$,$d$) combinations, we argue in the following
as for which conditions may be favourable for enhancing the efficiency
of soft THz emission. 

First, a mechanism is needed to create a sufficiently populated ``phonon bath'' from which
the excitation energy for electrons will be borrowed. As mentioned in Sec.~\ref{sec:Intro},
the GNP of sizes ${\simeq}\,5-7$~nm
(i.e., $N_x\,{\simeq}\,12-17$) have been reported to be efficiently heated by RF irradiation
(see \cite{NanoRes2-400}, also our interpretation of this effect in Ref.~\cite{FANEM2015}),
that seems quite promising in terms of applications. The skin effect depth in gold
is 25~nm at 8.7~THz (for comparison: 100~nm at 0.54~THz) \cite{SkinEffect_calc},
therefore the GNPs of sizes under discussion will be transparent for ``soft'' and
``hard'' generated THz radiation.
One can see from Fig.~\ref{fig:phonon}b that there are only few ``good'' matchings
in the sense of Eq.~(\ref{eq:matching}) for the above values of $N_x$;
however, one can presume that in practical conditions, a large array of GNPs deposited
on a substrate won't be very selective and would emit throughout a certain range
of ``soft THz'' frequencies. So, the GNPs preferential size is essentially chosen by
a tendency to yield a convenient ``pumping'' mechanism through RF irradiation.

The second consideration would be to ensure that the relaxation of an excited electron
proceeds through emitting a THz photon rather than, say, a soft longitudinal phonon.
In fact, due to the bending down of a phonon dispersion branch towards the BZ boundary
-- see Fig.~\ref{fig:phonon}a and Eq.~(\ref{eq:ha-nu}) -- the minimal phonon energy 
$h\nu(\zeta\!=\!1/N_x)$ exceeds by far the energy difference of phonons near the BZ
boundary, $h\nu(\zeta_1\!=\!1)\!-\!h\nu(\zeta_2\!=\!1\!-\!d/N_x)$ up to the numbers 
of quantisation steps $d=5-6$ throughout $N_x=11-20$. Consequently, for small
phonon momentum mismatch between the absorbed/emitted BZ-boundary phonons
($d=1-2$ steps) the excitation energy ${\delta}E$ cannot be channeled into a zone-center
acoustic phonon because the latter is ``too heavy'' to excite.
In case of larger mismatch $d$, the energy difference may eventually suffice for
emitting a small-$\mathbf{q}$ acoustic phonon; however, in this case it would be
impossible to satisfy the momentum conservation.

\section{Case of two-phonon absorption}
\label{sec:2ph-abso-abso}
The relevance of processes with two-phonon absorption follows
from an observation that the radius of the Fermi surface of gold
is only slightly shorter than the distance to the BZ boundary.
The free-electron model for a fcc-lattice monovalent metal  
with lattice parameter $a$ yields
\begin{equation}
k_{\rm F}=\frac{1}{a}\sqrt[3]{12\pi^2} = \frac{2\pi}{a}\!\left(
\!\frac{3}{2\pi}\!\right)^{\!\!\frac{1}{3}}
\approx \frac{2\pi}{a}\,{\cdot}\,0.782\,,
\label{eq:k_F_fcc}
\end{equation}
that is in fact a fairly good estimate in case of gold, whose Fermi radius measures,
in units of $2\pi/a$, ${\simeq}\,0.778$  along ${\langle}211{\rangle}$ and
${\simeq}\,0.737$ along ${\langle}110{\rangle}$ \cite{PRB25-7818}. 
As we compare this to the phonon dispersion $\nu(\zeta)$ in Fig.~\ref{fig:phonon}a
we readily see that $\zeta$ values $0.74\!-\!0.78$ yield $\nu\!=4.30\!-\!4.38$~THz,
that is within the ``useful'' range of the phonon spectrum
(within the major peak related to longitudinal phonons). 
An absorption of two such phonons would, in principle, promote an electron
to a nearly opposite spot of the Fermi surface. We argue in the following that,
for the particle sizes under discussion, the momentum conservation condition
can be ``absorbed'' by the uncertainty relation, thus enabling a certain tolerance
in the exact matching of quantified electron / phonon momenta.

The absorption of two phonons of $\simeq\,$4.4~THz frequency would mean 
exciting an electron far beyond the lowest quantisation level ${\delta}E$
of Eq.~(\ref{eq:delta-E}). In fact, setting ${\delta}E\,{\simeq}\,36$~meV
(that corresponds to $\nu\!=\,$8.7~THz) in Eq.~(\ref{eq:delta-E})
yields $N_x\,{\simeq}\,4-5$, or $L\,{\simeq}\,1.8$~nm, clearly too small
to justify our present ``condensed matter-like'' argumentation and, moreover,
unreasonably small for practical GNP applications we have in mind.
For GNP sizes discussed above, i.e., $N_x\,{\simeq}\,10-20$, the excitation energy
${\Delta}E$ on the absorption of two phonons with reduced wavenumbers $\zeta_1$,
$\zeta_2$ would comprise an integer number $m$ of minimal steps
${\delta}E$ from Eq.~(\ref{eq:delta-E}):
\begin{equation}
{\Delta}E = m\,{\delta}E = h\nu(\zeta_1)+h\nu(\zeta_2)\,.
\label{eq:Delta-E}
\end{equation}
From Eq.~(\ref{eq:delta-E}), $N_x$=10 would yield $m\,{\simeq}12$
($m\,{\simeq}\,16$ for $N_x$=11, $m\,{\simeq}\,21$ for $N_x$=12, etc.).
Because of the liberty in the choice of the matching $m$ value, presumably it won't be
difficult to find a pair of phonons whose energies would fit Eq.~(\ref{eq:Delta-E}).

Similar to Eq.~(\ref{eq:delta-pF}), i.e., replacing the energy dispersion
for electrons by its linear approximation throughout the ${\Delta}E$ range,
and independently of $N_x$, we get
${\Delta}p_{\rm F}\,\simeq\,(p_{\rm F}/2)(\Delta E_{\rm F}/E_{\rm F})\,\simeq\,
4.13{\cdot}10^{-22}$~g$\cdot$cm/s, i.e.,
still by a factor of $\simeq\,40$ smaller than the quantisation step
in the momentum of phonons, Eq.~(\ref{eq:delta-hq}), and by a factor of $\simeq\,300$
smaller than $p_{\rm F}$.

We turn now to more attentive consideration of the momentum conservation.
We note that the momentum of an electron in the confined space of linear size $L$
is subject to uncertainty (in the sense of standard deviation, taking $L/2$
for that concerning the position)
\begin{equation}
{\Delta}p \geq \frac{\hbar}{L} = \frac{1}{2\pi}\,\delta(\hbar q) \approx
0.16\,{\cdot}\,\delta(\hbar q)\,,
\end{equation}
$\delta(\hbar q)$ being the step in the momentum quantisation for phonons, Eq.~(\ref{eq:delta-hq}).
The uncertainty of ${\pm}0.16$ this value would certainly ``absorb'' the ``perturbation''
due to the electron excitation ${\Delta}p_{\rm F}$ and otherwise serve as a measure
of an ``exact'' matching. For the two absorbed phonons being ``at the distances'' $s_1$
and $s_2$ from the BZ boundary, i.e., $\zeta_1\!=\!1\!-\!s_1/N_x$ and $\zeta_2\!=\!1\!-\!s_2/N_x$,
the momentum conservation condition (for the case when the phonon momenta are parallel)
would read:
\begin{equation}
p_1+p_2 = \frac{2\pi}{a_{\rm Au}}\hbar\left(\!2-\frac{s_1\!+\!s_2}{N_x}\!\right)=
2p_{\rm F}+{\Delta}p_{\rm F} \approx 2p_{\rm F}\,;
\end{equation}
whence, by force of Eq.~(\ref{eq:k_F_fcc}), 
\begin{equation}
\frac{s_1\!+\!s_2}{N_x}=2\left[1-\left(\!\frac{3}{2\pi}\!\right)^{\!\!\frac{1}{3}}\right]
\approx 0.437\,.
\end{equation}
Searching for ``nearly integer'' (within about ${\pm}\,0.16$) solutions
for $(s_1\!+\!s_2)$,
one finds them the more easily the larger the particle size $N_x$, 
e.g., $(s_1\!+\!s_2)/N_x = 5/11;\; 6/14;\; 7/16;\; 8/18 \dots$

An \emph{a priori} imaginable process of absorption of two phonons with opposite momenta
would confront a problem of the smallness of ${\Delta}p_{\rm F}$. Indeed, the corresponding
matching condition for momenta would read
\begin{eqnarray}
p_2-p_1 &=& \frac{2\pi}{a_{\rm Au}}\,\hbar\,\frac{s_2\!-\!s_1}{N_x} = \Delta p_{\rm F}\,;
\nonumber \\
\frac{s_2\!-\!s_1}{N_x} &=& \frac{a_{\rm Au}}{h}\,\frac{p_{\rm F}}{2}\,
\frac{\Delta E_{\rm F}}{E_{\rm F}} = 
\frac{1}{2}\left(\!\frac{3}{2\pi}\!\right)^{\!\!\frac{1}{3}}\!
\frac{\Delta E_{\rm F}\!=\!36\mbox{~meV}}{E_{\rm F}\!=\!5.53\mbox{~eV}} 
\simeq 0.0025\,, 
\end{eqnarray}
``useless'' up to particle sizes $N_x\,{\gtrsim}\,400$.
 However, one cannot exclude 
a special situation $(s_1\!=\!s_2)$ when two phonons with exactly opposite momenta 
are absorbed and promote an electron to ${\Delta}E$, whereby the necessary excess momentum
${\Delta}p_{\rm F}$ is ``generated'' by the uncertainty of the phonon momenta.

Once the electron is excited via a two-phonon absorption, its relaxation 
cannot proceed via another channel than emission of a THz photon,
since the phonon spectrum of gold does not contain phonons of corresponding frequency. 

\section{Discussion on practical realisation}
\label{sec:discus}
In the above analysis, the reference was repeatedly made to
GNPs in the shape of rhombicuboctahedra, for the simple reason that 
such particles, grown from rhombic dodecahedral seeds
to the diameter of $\sim$35~nm, come about as intermediate stage
in the controllable growth / etching process towards either cubic
or octahedral particles of about double this size
\cite{CrystEngComm12-116,ChemEurJ17-9746}. 
In view of obvious distortion / surface relaxation / rounding at the edges
it hardly makes sense to refer to an ``exact'' dispersion relation
for phonons in the polyhedral particle serving as a ``resonator''.
Still, wall-to-wall confinement conditions between the delimiting (100) planes
of the rhombicuboctahedron seem to offer a clearly defined and realistic model.
Further factors, e.g., confinement of phonons between opposite (110)
or (111) faces of the rhombicuboctahedron, as well as irregularities of shape,
would result in smearing out of the sharp quantisation conditions. 

The ``interesting'' size of GNPs would be somehow smaller than the $\sim$35~nm
cited above; notably, as was already mentioned, the most efficient heating 
by RF irradiation seems to occur for the particles of ${\simeq}\,5-7$~nm size
\cite{NanoRes2-400}, or, in our counting, $N_x\,{\simeq}\,12-17$.
If our explanation \cite{FANEM2015} of this size selectivity is valid,
the essential mechanism of GNP heating, via phonon-assisted absorption
of electromagnetic quanta with energies much smaller than the electron
quantisation step ${\delta}E_{\rm F}$, would hold irrespectively
of the exact energy of the pumping field. Namely, one can envisage using
the ``pump frequency'' throughout the range from RF at 13.56~MHz ($h\nu=5.6{\cdot}10^{-5}$~meV)
as in Ref.~\cite{NanoRes2-400} up to the domestic microwave oven operating
at 2.45~GHz ($h\nu=1.0{\cdot}10^{-2}$~meV), since the respective $h\nu$ values remain by far
inferior to ${\delta}E$ of Eq.~(\ref{eq:delta-E}) which makes, e.g., 0.37~meV for $N_x=20$.

\begin{figure}[t]
\flushright{\includegraphics[width=0.85\textwidth]{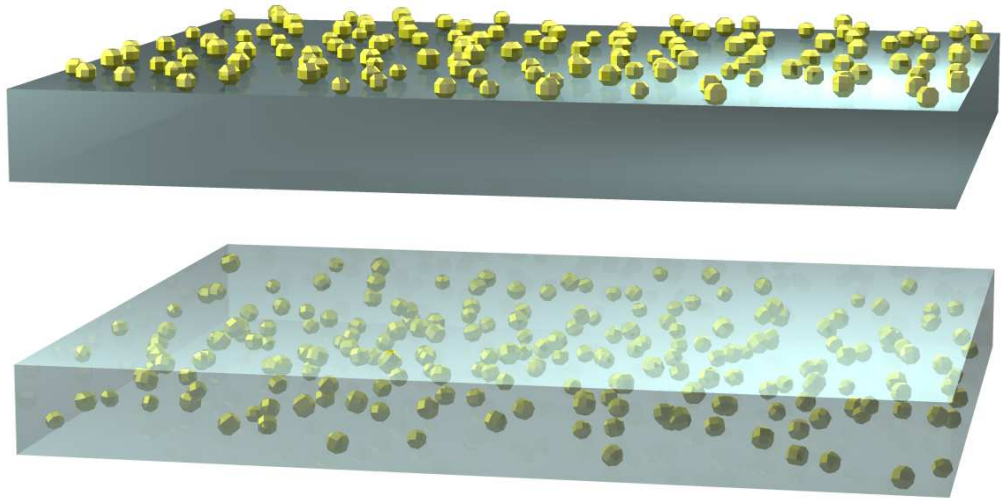}}
\caption{\label{fig:substrate}
Schematic views of an array of GNPs
in the shape of rhombicuboctahedra deposited on (the upper figure)
or within (the lower figure) a THz-transparent layer.
}
\end{figure}

In a practical design, the GNPs need to be fixed on/in some substrate
transparent for RF and THz irradiation (Fig.~\ref{fig:substrate}),
well isolating the generated heat
in order to channel the RF energy into phonon bath and then THz emission,
rather than dissipating into the substrate. 
The Teflon\textsuperscript{\textregistered} seem to be a convenient
material for this, as has been discussed in Ref.~\cite{Ferroel509-158},
citing its transmission properties, among other THz-relevant materials,
from Ref.~\cite{THz-Materials}.
Preparation methods for aqueous dispersions of polytetrafluoroethylene
have been described by patents 
\cite{Patent-RU2387672,Patent-US7141620B2,Patent-US7342066B2,Patent-US20080015304A1};
commercially produced gold nanoparticles suspended in water (colloidal gold) 
are also available, therefore the fabrication of  
Teflon\textsuperscript{\textregistered}-embedded GNPs seems feasible.
Their irradiation with standard sources (RF in the MHz range, or a microwave oven
operating at 2.45~GHz) would generate heating up to 260$^{\circ}$
(the maximal working temperature sustainable by Teflon)
and channel the energy of thus induced phonons into THz radiation. 
An advantage over existing THz sources is an absence of sophisticated 
or expensive active elements and a potentially large cross-section
of the THz beam, limited only by the size of the array of substrate-embedded
nanoparticles. Focusing the beam with THz lenses \cite{THz-Lenses}
would, if necessary, correspondingly enhance the THz power density.

In view of further enhancing the THz radiation yield, it could be useful to increase
the number of Fermi electrons within the nano-object. This can be achieved by using the metals
characterised by an elevated density of electronic states at the Fermi level,
e.g., Pd, Pt or Ni, or doping the GNPs with metal impurities known to place a pronounced
density of states peak at $E_{\rm F}$. Obviously, a simple free-electron model won't be
valid for estimating the parameters of such particles. 

Along with compact GNPs, gold nanobars with lengths up to micrometers
and transversal size of $\approx\,$5~nm can be presumably used for generating
THz radiation, via ``conversion'' of longitudinal phonons propagating
along the nanobar. The basic principles of such generation have been earlier
addressed in Ref.~\cite{BeilJNano7-983}.
Like in the case of compact GNPs, conditions can be created for two-phonon
involvement in such generation, either via absorption-emission of phonons,
accompanying an emission of ``soft'' (difference energy) THz photon, or via
two-phonon absorption resulting in ``hard'' (summary energy) THz photon emission.
The limiting cases and criteria for such processes will be discussed elsewhere.                                                                          

\section{Conclusion}
\label{sec:conclu}
Summarizing, the possibility of THz generation by GHz-irradiated arrays of
metallic nanoparticles deposited on (or, embedded into) dielectric 
(suggested: Teflon\textsuperscript{\textregistered}) substrate,
addressed in previous works, is now extended over the inspection of a possibility 
to emit ``soft THz''  (at $\sim$0.54~THz) and ``hard THz'' (at $\sim$8.7~THz)
radiation via two-phonon processes. We do not know how a relative outcome 
of such processes can be tuned; they seem simply to contribute to the total
emission yield. However, ``soft'' or ``hard'' ranges of the spectrum can be
selected by appropriate filtering \cite{THz-Filters}. The two ranges may have
specific (distinct) advantages, e.g., for medical screening of tumours,
or wound inspection under bandages.
The ``soft'' 0.54~THz range enhances the contrast in representing the normal
vs pathologic tissues, whereas the ``hard'' 8.7~THz range enhances the spatial
resolution, both radiation ranges being harmless for living objects.

\section*{References}

\end{document}